\documentclass[conference]{IEEEtran}
\IEEEoverridecommandlockouts
\usepackage{cite}
\usepackage{amsmath,amssymb,amsfonts}
\usepackage{algorithmic}
\usepackage{graphicx}
\usepackage{textcomp}
\usepackage{url} 
\usepackage{xcolor}
\usepackage{tikz}
\usetikzlibrary{shapes.geometric, arrows, positioning}
\usepackage{fontawesome5}  
\usepackage{amsthm}
\usepackage{subcaption}

\theoremstyle{plain} 
\newtheorem{theorem}{Theorem}
\newtheorem{corollary}{Corollary}

\def\BibTeX{{\rm B\kern-.05em{\sc i\kern-.025em b}\kern-.08em
    T\kern-.1667em\lower.7ex\hbox{E}\kern-.125emX}}
\begin{document}

\title{FORTA: Byzantine-Resilient FL Aggregation via DFT-Guided Krum}
\author{
Usayd Shahul$^{\dagger}$ and J.~Harshan$^{*, \dagger}$ \\
$^\dagger$Bharti School of Telecomunication Technology and Management, $^{*}$Department of Electrical Engineering,\\
Indian Institute of Technology Delhi, India
}
\maketitle


\begin{abstract}
Secure federated learning enables collaborative model training across decentralized users while preserving data privacy. A key component is secure aggregation, which keeps individual updates hidden from both the server and users, while also defending against Byzantine users who corrupt the aggregation. To this end, Jinhyun So et al. recently developed a Byzantine-resilient secure aggregation scheme using a secret-sharing strategy over finite-field arithmetic. However, such an approach can suffer from numerical errors and overflows when applied to real-valued model updates, motivating the need for secure aggregation methods that operate directly over the real domain.

We propose \textit{FORTA}, a Byzantine-resilient secure aggregation framework that operates entirely in the real domain. FORTA leverages Discrete Fourier Transform (DFT) codes for privacy and employs Krum-based outlier detection for robustness. While DFT decoder is error-free under infinite precision, finite precision introduces numerical perturbations that can distort distance estimates and allow malicious updates to evade detection. To address this, FORTA refines Krum using feedback from DFT decoder, improving the selection of trustworthy updates. Theoretical analysis and experiments show that our modification of Krum offers improved robustness and more accurate aggregation than standard Krum.
\end{abstract}


\section{Introduction}

Federated learning (FL) enables decentralized model training across distributed devices without collecting raw datasets centrally~\cite{mcmahan2017communication}. In this setting, a central server coordinates the process by broadcasting a global model, which each user trains locally before sending updates back for aggregation. This process continues until the global model stabilizes and achieves reasonable performance. Despite not sharing raw datasets, local updates can still leak sensitive information. In particular, adversaries can exploit gradient inversion techniques to reconstruct private training datasets~\cite{geiping2020inverting,zhu2019deep}. Moreover, FL is also vulnerable to \textit{Byzantine attacks}, where malicious users send corrupted updates to degrade model performance. As shown in~\cite{blanchard2017byzantine}, even a single malicious user can significantly skew aggregation under schemes like FedAvg~\cite{mcmahan2017communication}.
To defend against Byzantine faults, a common strategy is to compare local updates from different users and filter out anomalous ones at the server~\cite{blanchard2017byzantine,guerraoui2018hidden,yin2018byzantine}. However, these strategies require the server to access individual updates thereby posing a privacy risk.
Thus, designing FL systems that are both Byzantine-resilient and privacy-preserving presents a fundamental challenge, i.e.,  Byzantine resilience requires visibility into updates, while privacy demands that updates remain hidden. Therefore, reconciling these conflicting goals is nontrivial.

Recent works~\cite{so2020byzantine,jahani2023byzantine} have explored secure aggregation techniques that integrate Byzantine resilience while preserving  privacy. These methods typically rely on cryptographic primitives such as secret sharing and commitment schemes to enable privacy-preserving computations over masked model updates. The central idea is to allow the server to assess the reliability of user updates, often through pairwise distance-based anomaly detection, without directly accessing individual model parameters. By leveraging coded computing or homomorphic properties, these methods enable secure distance computation while maintaining robustness against adversarial manipulations.  Once malicious updates are identified as outliers, only the selected user updates contribute to the final model aggregation, ensuring both resilience to Byzantine attacks and privacy preservation. 
In particular, the approach presented in\cite{so2020byzantine} uses secret sharing over finite fields to enable secure distance computation. However, its effectiveness depends heavily on the choice of the field size. Accurate distance recovery requires a large field size, which introduces the following challenges:
First, a large field size leads to high computational overhead in modular arithmetic and interpolation, especially in large-scale FL settings. Second, since model updates are real-valued (floating point), quantizing them into a finite field can cause accuracy loss due to overflows and rounding errors.
Therefore, in practice, finite field based secure aggregation must trade off between a small field size which risks reconstruction errors and large field size that increases cost. These limitations hinder the scalability and accuracy of existing finite-field-based approaches, thereby opening up new alternatives in the real domain. 

\subsection{Contributions}

We propose \textit{Fourier-Based Outlier-Resilient Trust Aggregation (FORTA)}, a secure aggregation framework for FL that operates entirely in the real domain. Unlike schemes that rely on finite-field arithmetic, \textit{FORTA} eliminates quantization and enables direct aggregation of real-valued model updates. It is designed to: (i) preserve privacy, (ii) ensure robustness against adversarial updates, and (iii) remain effective under finite precision constraints.

A key challenge in FL is defending against \textit{model poisoning attacks}, where Byzantine users inject malicious updates. \textit{FORTA} addresses this by integrating the \textit{Krum} aggregation rule~\cite{blanchard2017byzantine}, which identifies non-malicious users based on \textit{pairwise distances} between updates. To preserve privacy, it leverages \textit{Analog secret sharing}~\cite{soleymani2020privacy}, enabling secure distance computation without exposing raw updates.
However, Byzantine users can corrupt secret shares to distort distances as well. To correct this, \textit{FORTA} uses Discrete Fourier Transform (DFT)-based error correction algorithm. Yet, real-domain decoding is vulnerable to \textit{finite precision errors}, which adversaries can exploit to evade detection. To overcome this, we incorporate joint decoding and  feedback from the decoder to enhance robustness.
 Our main contributions are:
\begin{enumerate}
    \item \textbf{Real-domain secure aggregation with Krum:}  
    We introduce a real-domain secure aggregation framework combining analog secret sharing with Krum-based outlier detection, enabling privacy-preserving pairwise distance computation without quantization.

    \item \textbf{GMM-based error localization for finite precision:}  
    We propose a joint decoding strategy based on Gaussian Mixture Models (GMMs) that improves adversarial localization by distinguishing structured attacks from precision noise in the DFT decoder.

    \item \textbf{Subtle attack via finite precision exploitation:}  
     Despite combining Krum with GMM-based decoding, we identify a novel attack strategy where Byzantine users craft perturbations that closely mimic finite-precision noise. This exploits imperfections in DFT decoder accuracy to bypass distance-based outlier detection and compromise secure aggregation.

    \item \textbf{Modified Krum with DFT-guided feedback:}  
      To counter this vulnerability, \textit{FORTA} introduces a modified Krum rule that integrates frequency-based feedback from the DFT decoder. This enhances the selection of reliable users, significantly improving robustness under precision-aware Byzantine attacks.

    \item \textbf{Theoretical and empirical validation:}  
    We provide theoretical resilience guarantees for our modified Krum and validate its performance on MNIST and CIFAR-10, showing improved robustness under Byzantine attacks.
\end{enumerate}

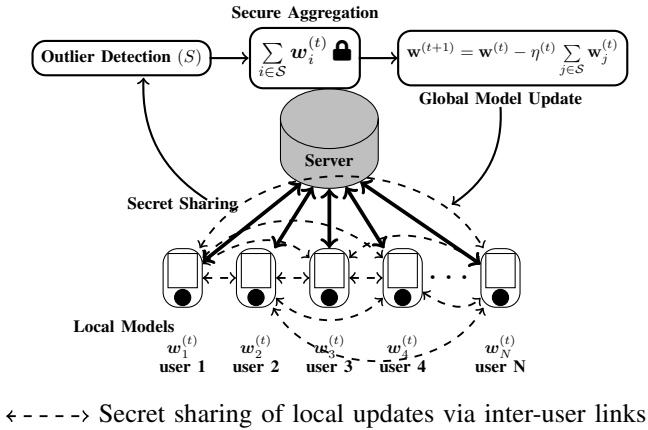
\begin{figure}[htbp] 
\begin{center}
\begin{tikzpicture} [transform shape, scale=0.65]

     \node[draw, cylinder, shape border rotate=90, minimum width=2cm, minimum height=1.5cm, fill=lightgray] 
        (server) at (0,2.4) {\textbf{Server}};
    
    
    
    \node[draw, thick, rounded corners=5pt, inner sep=5pt] 
    at (-4.25,4.5) {\normalsize \textbf{Outlier Detection }$\mathcal({S})$};

\draw[->, thick] (-2.45,4.5) -- (-1.65,4.5); 

\node[draw, thick, rounded corners=5pt, inner sep=5pt] 
    at (-0.5,4.5) {\scalebox{1.2}{$\sum\limits_{i\in \mathcal{S}} \boldsymbol{w}_i^{(t)}$} \hspace{-4pt} \scalebox{1.2}{\faLock}};

\draw[->, thick] (0.6,4.5) -- (1.4,4.5); 

\node[draw, thick, rounded corners=5pt, inner sep=5pt, fill=white] 
    at (3.75,4.5) {\normalsize $\mathbf{w}^{(t+1)} = \mathbf{w}^{(t)} - \eta^{(t)} \sum\limits_{j \in \mathcal{S}} \mathbf{w}_j^{(t)}$};

    \foreach \angle/\name/\pos/\index in {150/user 1/-3/1, 110/user 2/-1.5/2, 70/user 3/0/3, 30/user 4/1.5/4} {
        \node[draw, rectangle, minimum width=0.8cm, minimum height=1.2cm, rounded corners=0.2cm] 
            (\name) at (\pos,0) {};
        \node[draw, rectangle, fill=white, minimum width=0.6cm, minimum height=0.7cm] 
            at (\pos,0.15) {};
        \node[draw, circle, fill=black, minimum size=0.15cm] 
            at (\pos,-0.4) {};    
        \node[font=\bfseries] at (\pos,-1.8) {\name};
        \node[font=\bfseries] at (\pos, -1.4) {$\boldsymbol{w}_{\index}^{(t)}$};
    }
    
    \node[font=\bfseries] at (2.5,0) {\huge $\cdots$};

    \node[draw, rectangle, minimum width=0.8cm, minimum height=1.2cm, rounded corners=0.2cm] 
        (user N) at (3.5,0) {};
    \node[draw, rectangle, fill=white, minimum width=0.6cm, minimum height=0.7cm] 
        at (3.5,0.15) {};
    \node[draw, circle, fill=black, minimum size=0.15cm] 
        at (3.5,-0.4) {};    
    \node[font=\bfseries] at (3.5,-1.8) {user N};
    \node[font=\bfseries] at (3.5, -1.4) {$\boldsymbol{w}_{N}^{(t)}$};
    \foreach \i in {1,2,3,4,N} {
        \draw[line width =0.5mm, <->] (user \i) -- (server);
    }
\draw[dashed, <->, line width=0.25mm] (user 1) to (user 2);
\draw[dashed, <->, line width=0.25mm] (user 2) to (user 3);
\draw[dashed, <->, line width=0.25mm] (user 3) to (user 4);

\draw[dashed, <->, line width=0.25mm, out=35, in=135] (user 1) to (user 3);
\draw[dashed, <->, line width=0.25mm, out=45, in=135] (user 1) to (user 4);
\draw[dashed, <->, line width=0.25mm, out=60, in=120] (user 1) to (user N);
\draw[dashed, <->, line width=0.25mm, out=-45, in=-135] (user 2) to (user 4);
\draw[dashed, <->, line width=0.25mm, out=-60, in=-120] (user 2) to (user N);
\draw[dashed, <->, line width=0.25mm, out=45, in=145] (user 3) to (user N);
\draw[dashed, <->, line width=0.25mm, out=-35, in=-135] (user 4) to (user N);

\draw[->, thick, bend left=30] (-2.5,1.6) to (-3.8,4.1);
\draw[->, thick, bend left=30] (3.5,3.5) to (2.5,1.5);

    \node[align=left, font=\bfseries\normalsize] at (-3,1.5) {\textbf{Secret Sharing}};
\node[align=left, font=\bfseries\normalsize] at (-4.2,-1) {\textbf{Local Models}};
\node[align=center, font=\bfseries\normalsize] at (-0.5,5.4) {\textbf{Secure Aggregation}};
\node[align=left, font=\bfseries\normalsize] at (3.5,3.65) {\textbf{Global Model Update}};
\end{tikzpicture}
\end{center}
\begin{center}
\begin{tikzpicture}
  
    \draw[dashed, <->, line width=0.25mm] (0,-0.7) -- (1.1,-0.7);
    \node[right] at (1.11,-0.7) {Secret sharing of local updates via inter-user links};
\end{tikzpicture}
\end{center}

\caption[Diagram of Federated Learning Framework]{Secure FL framework with outlier detection.}
\label{fig:FL-diagram}  
\end{figure}

\section{System Overview}

We consider an FL setup with \( N \) users and a central server. Each user \( i \in [N] \) holds a private dataset \( \mathcal{D}_i \) and can communicate with the server (user-to-server link) and with other users (user-to-user link) in a fully connected topology, as shown in Fig.~\ref{fig:FL-diagram}. The goal is to collaboratively train a global model \( \boldsymbol{w}^{(t)} \in \mathbb{R}^d \) at iteration \( t \geq 0 \) that minimizes the global loss $\mathcal{L}(\mathbf{w}) = \displaystyle\frac{1}{N} \sum_{i=1}^{N} n_i \mathcal{L}_i(\mathbf{w})$, where \( \mathcal{L}_i(\mathbf{w}) \) is the local loss over \( \mathcal{D}_i \).
The training proceeds in rounds as follows:

\begin{itemize}
    \item \textbf{Broadcast:} The server sends the current global model \( \boldsymbol{w}^{(t)} \) to all users.

    \item \textbf{Local Update:} Upon receiving the global model \( \boldsymbol{w}^{(t)} \), each user \( i \in [N] \) computes local update as  $\boldsymbol{w}_i^{(t)} = g(\boldsymbol{w}^{(t)}, \xi_i^{(t)})$, where \( g\) is an estimate of the gradient \( \nabla\mathcal{L} (\boldsymbol{w}^{(t)}) \) of the loss function \( \mathcal{L} \), and \( \xi_i^{(t)} \) denotes the mini-batch of data sampled by user \( i \) in iteration \( t \), and the stochastic gradient satisfies $
        \mathbb{E}_{\xi_i^{(t)}} [g(\boldsymbol{w}^{(t)}, \xi_i^{(t)})] = \nabla \mathcal{L}_i(\boldsymbol{w}^{(t)}).$
  
    \item \textbf{Aggregation:} The server performs secure aggregation with outlier detection to filter unreliable updates and refine the model as  
$\boldsymbol{w}^{(t+1)} = \boldsymbol{w}^{(t)} - \eta^{(t)} f(\boldsymbol{w}_1^{(t)}, \dots, \boldsymbol{w}_N^{(t)}),$  
where \( \eta^{(t)} \) is the learning rate and \( f \) is a robust aggregation rule over the filtered set.

\end{itemize}

Training continues iteratively until the global model reaches \textit{convergence}, implying that the model stabilizes with minimal changes in subsequent iterations.

\subsection{Threat Model}
Our FL setup involves up to \( A \) Byzantine users which are malicious participants aiming to disrupt model training process. These users may launch \textit{model poisoning attacks} by submitting corrupted updates to bias the global model and degrade performance. More subtly, they may inject  controlled perturbations during the protocol execution, aiming to corrupt
the aggregated results or delay the learning process.
We also assume an \textit{honest-but-curious server} that follows the protocol but seeks to infer private information from user updates. Additionally, up to \( T \) users may \textit{collude}, pooling private information in attempts to reconstruct others’ local updates.
This combination of threats namely, poisoned updates, protocol-level attacks, privacy violations, and user collusion can severely compromise model accuracy and convergence. A robust defense must therefore ensure privacy preservation, detect outliers, and remain resilient under finite precision errors.
We now proceed to formally describe the technical details of \textit{FORTA}.

\section{FORTA}
In each round \( t \) of FL, the FORTA framework performs a three-step pipeline that ensures privacy-preserving and robust aggregation in the presence of Byzantine users. These steps, repeated until convergence of the global model, consist of:  secret sharing of local updates over infinite fields,  outlier detection with feedback from the DFT decoder, and  secure aggregation of selected users followed by a global model update. We now describe each component in detail.

\subsection{Secret Sharing of Local Updates over Infinite Fields}
Our approach, inspired by analog secret sharing \cite{soleymani2020privacy}, operates in the real domain, encoding updates into polynomials with additive Gaussian noise. This enables privacy preservation without requiring discretization. However, in finite precision settings, numerical errors inevitably arise due to floating-point operations, introducing an additional noise term.

Let user \( i \) hold a local model update \( \boldsymbol{w}_i^{(t)} \in \mathbb{R}^d \). For notational simplicity, we omit the superscript \( (t) \) in \( \boldsymbol{w}_i^{(t)} \) throughout the rest of the paper. Initially, the server and users agree on \( N \) distinct elements \(\{\omega_i\}_{i \in [N]}\), where each \( \omega_i = e^{2\pi \iota i / N} \), and \( \iota \) denotes the imaginary unit with \( \iota^2 = -1 \). Thus, each \( \omega_i \) is an \( N^{\text{th}} \) root of unity in the complex plane. These points serve as evaluation points for the secret-sharing polynomials.  To ensure privacy, user \( i \) constructs a secret-sharing polynomial \( {P}_i: \mathbb{C} \to \mathbb{C}^d \) of degree \( T \) in indeterminate \(x\), given by:
\begin{equation}
    {P}_i(x) = \boldsymbol{w}_i + \sum_{j=1}^{T} \boldsymbol{r}_{ij} x^j + \boldsymbol{\epsilon}_i,
\end{equation}
where \( T \) represents the maximum number of colluding users, \( \boldsymbol{w}_i \in \mathbb{R}^d \) is the user's local model update, \( \boldsymbol{r}_{ij} \in \mathbb{C}^d \) are {independent and identically distributed (i.i.d.) random vectors}, where each individual entry of \( \boldsymbol{r}_{ij} \) is also i.i.d. The entries of \( \boldsymbol{r}_{ij} \) are drawn from a circularly symmetric complex Gaussian distribution with zero mean and variance \( \sigma_n^2 / T \)
    $
        r_{ij}^{(k)} \sim \mathcal{N}(0, \sigma_{n}^2 / T), \quad \forall k \in \{1, \dots, d\},$
\(\boldsymbol{\epsilon}_i \in \mathbb{R}^d\) represents the finite precision noise, which arises due to numerical errors and follows an arbitrary distribution.
    
To securely distribute the secret shares of \( \boldsymbol{w}_i \),  user \(i\) evaluates  polynomial \( P_i(x) \) at the agreed-upon roots of unity. The corresponding secret share that user 
\(i\) sends to user 
\(j\) is given by:
\begin{equation}
    \boldsymbol{s}_{ij} = P_i(\omega_j).
\end{equation}  
This ensures that no subset of fewer than \( T + 1 \) users can reconstruct  \( \boldsymbol{w}_i \).
While the secret-sharing mechanism ensures that local updates remain private, it does not prevent Byzantine users from submitting \textit{maliciously altered updates}. Since the server does not directly observe individual updates, it must rely on an additional  mechanism to detect and mitigate such Byzantine behavior. To achieve this, we employ an \textit{outlier detection mechanism} based on pairwise distances between model updates, specifically leveraging the \textit{Krum} aggregation rule \cite{blanchard2017byzantine}. To enable distance computation, each user locally computes differences between the secret shares they receive. These masked differences act as \textit{proxies} for the true pairwise differences and are sent to the server for reconstruction. We describe this process in detail below.  

\subsubsection{Pairwise Difference Computation by Users}  
User \( i \) computes  pairwise differences between the secret shares received from all user pairs \( j, k \in [N] \) as:
\begin{equation}
    \boldsymbol{d}_{jk}^{i} = \boldsymbol{s}_{ji} - \boldsymbol{s}_{ki}.
\end{equation}
These differences are sent to the server by every user. Since the computation is performed on secret shares, user \( i \) does not learn the actual updates \( \boldsymbol{w}_j \) or \( \boldsymbol{w}_k \), thus preserving privacy.

\subsubsection{Reconstructing True Pairwise Differences at the Server}  
Upon receiving these  pairwise differences $\{ \boldsymbol{d}_{jk}^{i} \}$
 from all users, the server collects \( N \) such evaluations for each pair \( (j,k) \), forming a structured sequence of vectors. However, since Byzantine
users may send corrupted vectors, the server must correct those
errors before reconstruction. To ensure correctness, the server uses the polynomial structure of the pairwise differences of secret shares, which allows  to exploit redundancy for error correction. Specifically, we express the difference of secret sharing polynomials as:  
\begin{align*}
    h_{jk}(x) &= P_j(x) - P_k(x) \\
              &= (\boldsymbol{w}_j - \boldsymbol{w}_k) 
              + \sum_{t=1}^{T} (\boldsymbol{r}_{jt} - \boldsymbol{r}_{kt}) x^t 
              + (\boldsymbol{\epsilon}_j - \boldsymbol{\epsilon}_k),
\end{align*}
where \( h_{jk}(x) \in \mathbb{C}^d \) is a vector-valued polynomial of degree at most \( T \), and each evaluation \( h_{jk}(\omega_i) \in \mathbb{C}^d \) represents the masked pairwise difference sent by user \( i \).
The server thus receives \( N \) such evaluations for each pair \( (j, k) \), forming a matrix $
\boldsymbol{C}_{jk} = \big[ h_{jk}(\omega_1), h_{jk}(\omega_2), \dots, h_{jk}(\omega_N) \big]$,
where \( \boldsymbol{C}_{jk} \in \mathbb{C}^{d \times N} \). Each row of this matrix corresponds to a scalar-valued codeword of length \( N \), i.e., for each coordinate \( l \in \{1, \dots, d\} \), the server can extract the codeword
$
\boldsymbol{c}_{jk}^{(l)} = \big( h_{jk}^{(l)}(\omega_1),\ h_{jk}^{(l)}(\omega_2),\ \dots,\ h_{jk}^{(l)}(\omega_N) \big),
$
where \( h_{jk}^{(l)}(x) \) denotes the \( l \)-th coordinate of the vector-valued polynomial \( h_{jk}(x) \).

By considering the coefficients of \( h_{jk}^{l}(x) \) as message symbols and viewing its evaluations at the \( N \)-th roots of unity as a codeword, the server essentially obtains a noisy codeword from an \( (N, T+1) \) DFT code.
 Since there are \( \binom{N}{2} \) unique distance pairs and \( d \) coordinates, 
 the server is tasked with decoding a total of \( \binom{N}{2} \ d \) scalar codewords from an \( (N, T+1) \) DFT code. These codewords may contain adversarially corrupted entries and are further affected by finite precision noise.
 
To decode each of them, the server applies a DFT decoder to every scalar codeword \( \boldsymbol{c}_{jk}^{(l)} \) independently. This yields the coefficients of the corresponding scalar message polynomial \( h_{jk}^{(l)}(x) \), from which the server evaluates the polynomial at \( x = 0 \) to recover the \( l \)-th component of the pairwise difference. Aggregating these decoded components server reconstructs the full pairwise difference vector as
    $h_{jk}(0) = {\boldsymbol{w}}_j - {\boldsymbol{w}}_k + \boldsymbol{\epsilon_{jk}}$, from which it obtains  
    $
    d_{jk} := \lVert {\boldsymbol{w}}_j - {\boldsymbol{w}}_k + \boldsymbol{\epsilon_{jk}} \rVert^2,$
where, \(\boldsymbol{\varepsilon_{jk}}\) denotes the decoding error arising from finite precision arithmetic. 

In practice, the effectiveness of DFT-based reconstruction and specifically the ability to localize and correct adversarial corruptions relies critically on estimating the number of errors on each codeword. However, in finite precision settings, independent decoding of codewords becomes unreliable due to precision noise in the error localization step, allowing carefully crafted adversarial perturbations to evade detection.
Prior work~\cite{borah2024securing} shows that joint error localization, leveraging statistical patterns across multiple codewords significantly improves detection accuracy in such environments. Building on this insight, we develop a joint decoding strategy based on Gaussian Mixture Model (GMM) clustering, which isolates adversarial indices by statistically separating them from noise-affected but honest evaluations. The output is a frequency profile, denoted by $\mathbf{f}$, of length N, that tracks how often each user is flagged as adversarial across all codewords. This frequency profile is used both to localize errors during decoding and to guide the outlier detection step. Full technical details will be provided in the extended journal version of this work.

\subsection{Outlier Detection with feedback from DFT decoder}

The Krum algorithm~\cite{blanchard2017byzantine} assigns each user \( i \) a score
$
S_i = \sum_{j \in \mathcal{M}_i} \| \boldsymbol{w}_i - \boldsymbol{w}_j \|^2,
$
where \( \mathcal{M}_i \subset [N]\setminus\{i\} \) denotes the \( N - A - 2 \) users whose updates are closest (in Euclidean distance) to \( \boldsymbol{w}_i \). This score reflects how well \( \boldsymbol{w}_i \) aligns with the majority.
Krum then selects the  top \( m \) users  with the lowest scores, assuming non malicious users are mutually closer than Byzantine users.
However, under finite precision, both adversarial perturbations and precision noise can distort scores, causing Byzantine users to appear benign. To mitigate this, we incorporate feedback from the DFT decoder’s frequency profile built from joint GMM-based decoding across all codewords to assess adversarial likelihood.

\paragraph{Error Localization Feedback}
The frequency profile yields a vector $\mathbf{f}$ where $f_i$ denotes how often user $i$ is flagged as adversarial. We convert this into soft confidences via softmax function as $\lambda_i = \frac{\exp(f_i / \tau)}{\sum_{j=1}^N \exp(f_j / \tau)}$,
with temperature $\tau$ controlling sharpness. These soft confidences are then used to modulate the original Krum scores, biasing the selection against frequently flagged users.

\paragraph{Modified Scoring}
We define a stabilized baseline score $S_{\min} := \min_{1 \leq i \leq N} S_{i} / (N - A - 2)$, and compute the modified Krum score:
\begin{equation}
S_i^{\text{mod}} := \lambda_i S_i + (1 - \lambda_i) S_{\min}.
\end{equation}
This balances geometric consistency from Krum and statistical suspicion from the DFT decoder as a result users frequently flagged as adversarial retain higher influence from $S_i$, while less suspicious ones are nudged closer to $S_{\min}$.

\paragraph{Selection}
The server then selects the \( m \) users with the smallest $S_i^{\text{mod}}$ for secure aggregation.
\subsection{Secure Aggregation of Selected Users}

In the final phase, the server securely aggregates updates from the selected users, denoted by $\mathcal{U} \subset [N]$
 without the knowledge of individual updates. The server broadcasts \( \mathcal{U} \), and each user \( i \) computes the sum of secret shares received from users in \( \mathcal{U} \) as
$\boldsymbol{s}_{i} = \sum_{j \in \mathcal{U}} \boldsymbol{s}_{ji},$
and sends \( \boldsymbol{s}_i \) to the server. From these, the server reconstructs the aggregated model update \( \sum_{j \in \mathcal{U}} \boldsymbol{w}_j \) via a DFT decoder as discussed before.
Then the global model is then updated as:
$\boldsymbol{w}^{(t+1)} = \boldsymbol{w}^{(t)} - \eta^{(t)} \sum_{j \in \mathcal{U}} \boldsymbol{w}_j^{(t)}.$
This ensures robust and privacy-preserving aggregation, leveraging DFT decoder along with modified Krum.

\section{robustness guarantees}

In this section, we formally analyze the robustness of our proposed user selection strategy that integrates error localization feedback with the classical Krum rule. Our goal is to theoretically demonstrate how incorporating feedback from the DFT decoder improves resilience against Byzantine users. To facilitate this demonstration, we begin by recollecting the definition of robustness given in \cite{blanchard2017byzantine}, commonly referred to as \((\alpha, A)\)-Byzantine Resilience which quantifies the deviation between the  output of the selection rule (in our case, the modified Krum) and the true gradient vector in the presence of up to \( A \) Byzantine users.

\textbf{Definition 1 (\( \alpha, A \)-Byzantine Resilience)}  
Let \( 0 \leq \alpha < \frac{\pi}{2} \) and \( 0 \leq A \leq N \). Consider i.i.d. vectors \( \boldsymbol{w}_1, \dots, \boldsymbol{w}_N \in \mathbb{R}^d \), with \( \boldsymbol{w}_i \sim G \) representing true local updates, such that \( \mathbb{E}[G] = \mathbf{g} \) and \( \mathbb{E}[\| G - \mathbf{g} \|^2] = d \sigma_g^2 \), which includes arbitrary (possibly adversarial) vectors \( \boldsymbol{b}_1, \dots, \boldsymbol{b}_A \). A function \( f \) is said to be \((\alpha, A)\)-Byzantine resilient if, for any set of Byzantine indices \( \{j_1, \dots, j_A\} \), $
    f(\boldsymbol{w}_1, \dots, \boldsymbol{b}_{j_1}, \dots, \boldsymbol{b}_{j_A}, \dots, \boldsymbol{w}_N)$
satisfies the inequality
    $\mathbf{g}^\top \mathbb{E}[f] \geq (1 - \sin \alpha) \| \mathbf{g} \|^2$,
where \( \mathbf{g} \) is the expected true update direction. This ensures that the output produced by \( f \) remains sufficiently aligned with the true update, even in the presence of adversarial interference.
In our scenario, however, the server does not directly access the local model updates \( \boldsymbol{w}_1, \dots, \boldsymbol{w}_N \in \mathbb{R}^d \). Instead, it operates on noisy pairwise differences of the form $
    \boldsymbol{\Delta}_{jk} := \boldsymbol{w}_j - \boldsymbol{w}_k + \boldsymbol{\varepsilon}_{jk}, \quad \forall j \neq k$
where \( \boldsymbol{\varepsilon}_{jk} \in \mathbb{R}^d \) denotes additive noise arising from finite-precision arithemtic. To reconcile this setting with the assumptions of the above definition, we conceptually reinterpret these differences to define \textit{surrogate updates} that stand in for  local updates in the theoretical analysis. These adjusted updates preserve the key statistical properties required for the robustness guarantee, as described next.

We assume that each \( \boldsymbol{\varepsilon}_{jk} \) is an i.i.d. \( d \)-dimensional vector such that \( \boldsymbol{\varepsilon}_{jk} \sim F \) with $
    \mathbb{E}[F] = \boldsymbol{0}, \quad \mathbb{E}[\| F \|^2] = d \sigma_\varepsilon^2.$
To align with the theoretical framework of the original Krum analysis, which assumes access to (possibly perturbed) individual updates, we reinterpret each noisy pairwise difference, 
$\boldsymbol{\Delta}_{jk} = \boldsymbol{w}_j - \boldsymbol{w}_k + \boldsymbol{\varepsilon}_{jk}$
as arising from two independently perturbed updates, i.e., $\boldsymbol{w}_j + \boldsymbol{\varepsilon}_{jk}^1 \quad \text{and} \quad \boldsymbol{w}_k - \boldsymbol{\varepsilon}_{jk}^2,$ where we model the pairwise noise \( \boldsymbol{\varepsilon}_{jk} \sim \mathcal{N}(\boldsymbol{0}, \sigma_\varepsilon^2 \boldsymbol{I}_d) \) as the sum of two i.i.d. Gaussian vectors, i.e., $\boldsymbol{\varepsilon}_{jk} = \boldsymbol{\varepsilon}_{jk}^1 + \boldsymbol{\varepsilon}_{jk}^2, \quad \text{with } \boldsymbol{\varepsilon}_{jk}^1, \boldsymbol{\varepsilon}_{jk}^2 \sim \mathcal{N}(\boldsymbol{0}, \tfrac{1}{2} \sigma_\varepsilon^2 \boldsymbol{I}_d).$
This decomposition ensures that the total noise variance in the pairwise difference remains \( \sigma_\varepsilon^2 \), while enabling the construction of \emph{surrogate updates} of the form $\boldsymbol{v}_j := \boldsymbol{w}_j + \boldsymbol{\varepsilon}_{jk}^1, \quad 
\boldsymbol{v}_k := \boldsymbol{w}_k - \boldsymbol{\varepsilon}_{jk}^2.$ Then, each \( \boldsymbol{v}_i \in \mathbb{R}^d \) satisfies \( \mathbb{E}[\boldsymbol{v}_i] = \boldsymbol{g} \) and $
    \mathrm{Var}(\boldsymbol{v}_i) = d(\sigma_g^2 + \tfrac{1}{2} \sigma_\varepsilon^2).$ 
While the entries of each \( \boldsymbol{v}_i \) remain independent, corresponding coordinates across different vectors become correlated due to shared noise terms across pairwise interactions, i.e., $
    \mathrm{Corr}(\boldsymbol{v}_j^{(k)}, \boldsymbol{v}_k^{(k)}) > 0 \quad \text{for some } j \neq k.$
This deviation from independence necessitates a careful reinterpretation of robustness guarantees. Nevertheless, we show that resilience can still be established under this noise-injected model by appropriately bounding the second moment of the adjusted updates. We now proceed to formalize resilience guarantees of Krum under noisy surrogate updates.
\begin{theorem}
Assume that \( 2A + 2 < N \) and the following inequality holds
    $2\eta(N, A) \sqrt{d} \, \sigma' < \| {g} \|$,
where $
    \eta(N, A) = \sqrt{ \left( N - A + \frac{A(N - A - 2) + A^2 (N - A - 1)}{N - 2A - 2} \right)}$,
and \( \sigma' = \sqrt{\sigma_g^2 + \tfrac{1}{2} \sigma_\varepsilon^2}. \)
Let \( \boldsymbol{v}_1, \dots, \boldsymbol{v}_N \in \mathbb{R}^d \) denote the surrogate updates computed from noisy pairwise differences, such that each \( \boldsymbol{v}_i \) has i.i.d. entries with $
    \mathbb{E}_{G,F}[\boldsymbol{v}_i] = \mathbf{g}, \quad 
    \mathbb{E}_{G,F} \left[ \| \boldsymbol{v}_i - \mathbf{g} \|^2 \right] = d \left( \sigma_g^2 + \tfrac{1}{2} \sigma_\varepsilon^2 \right)$, 
but corresponding coordinates across different vectors are correlated, i.e.,  $
    \mathrm{Corr}(\boldsymbol{v}_i^{(k)}, \boldsymbol{v}_j^{(k)}) = \rho_{ij}^{k}, \quad \forall i \neq j,$
where
$\rho_{ij}^k = \frac{\operatorname{Cov}\left( \boldsymbol{v}_i^{(k)}, \boldsymbol{v}_j^{(k)} \right)}{\sigma'^2}.$
Then, the Krum function \( {f}_{kr} \) is \((\alpha, A)\)-Byzantine resilient with
\begin{equation}
    \sin \alpha = \frac{2\eta(N, A) \sqrt{d} \, \sigma'}{\|{\mathbf{g}} \|}, \quad \text{where } 0 \leq \alpha < \frac{\pi}{2}.
\end{equation}

\end{theorem}

We now state a similar  result for the {modified Krum} function. To formalize this result, we begin by introducing the relevant quantities and assumptions. Let \( \lambda_i  \in [0,1] \) denote a decoder-derived confidence  indicating the likelihood of user \( i \) being Byzantine. Define random variables
$T \sim \max \left( \frac{\lambda_i}{\lambda_k} \right), \quad Q \sim \max \left( 1 - \frac{\lambda_i}{\lambda_k} \right)$,
with corresponding means \( \mu_T, \mu_Q \) and variances \( \sigma_T^2, \sigma_Q^2 \).
The modified Krum score is given by $
S_i^{\text{mod}} = \lambda_i S_i + (1 - \lambda_i) \frac{S_0}{N - A - 2},$
where \( S_i = \sum_{j \to i} \| \boldsymbol{v}_i - \boldsymbol{v}_j \|^2 \), and \( S_0 = \min_{j \in S} S_j \) is the minimum original score. The notation \( j \to i \) indicates that \( \boldsymbol{v}_j \) is among the \( N - A - 2 \) closest vectors to \( \boldsymbol{v}_i \), and \( S \) denotes the set of original (unweighted) scores.

\begin{theorem}
Assume  \( 2A + 2 < N \) and  \( \boldsymbol{v}_1, \dots, \boldsymbol{v}_N \in \mathbb{R}^d \) denote the surrogate updates computed from noisy pairwise differences, such that each \( \boldsymbol{v}_i \) has i.i.d. entries with
 $
    \mathbb{E}_{G,F}[\boldsymbol{v}_i] = \mathbf{g}, \quad 
    \mathbb{E}_{G,F} \left[ \| \boldsymbol{v}_i - \mathbf{g} \|^2 \right] = d \left( \sigma_g^2 + \tfrac{1}{2} \sigma_\varepsilon^2 \right)=d\sigma'^2,$ 
but corresponding coordinates across different vectors are correlated, i.e.,  $
    \mathrm{Corr}(\boldsymbol{v}_i^{(k)}, \boldsymbol{v}_j^{(k)}) = \rho_{ij}^{k}, \quad \forall i \neq j,$
where $
\rho_{ij}^k = \frac{\operatorname{Cov}\left( \boldsymbol{v}_i^{(k)}, \boldsymbol{v}_j^{(k)} \right)}{\sigma'^2}.$
Assume further that there exist finite constant \( C_1 > 0 \) such that the second moment of each squared norm term satisfies
$
\mathbb{E} \left[ \left\| \boldsymbol{v}_i - \boldsymbol{v}_j \right\|^4 \right] 
\leq C_1 \cdot \left( \mathbb{E} \left[ \left\| \boldsymbol{v}_i - \boldsymbol{v}_j \right\|^2 \right] \right)^2.
\quad $

If the following condition holds
    $\sqrt{A(\eta'\Psi_T +\frac{\eta'}{N - A - 2} \Psi_Q)4d\sigma'^2  + (N - A)4d\sigma'^2} < \| \boldsymbol{g} \|$,
where
$
\eta' = \frac{N - A - 2}{N - 2A - 2} + \frac{A(N - A - 1)}{N - 2A - 2}, \quad
\Psi_T = \sqrt{ (\sigma_T^2 + \mu_T^2) \cdot C_1 }, \quad
\Psi_Q = \sqrt{ (\sigma_Q^2 + \mu_Q^2) \cdot C_1 }$,
then  the modified Krum function \( {f}_{\text{mod-kr}} \) is \( (\alpha', A) \)-Byzantine resilient, where
\begin{equation}
      \sin \alpha' = \frac{\sqrt{A(\eta'\Psi_T +\frac{\eta'}{N - A - 2} \Psi_Q)4d\sigma'^2  + (N - A)4d\sigma'^2 }}{\| \boldsymbol{g} \|}.
\end{equation}
\end{theorem}

\begin{corollary}
Let \( \alpha \) and \( \alpha' \) denote the deviation angles associated with the original and modified Krum functions as stated in Theorems~1 and~2, respectively. 
If the following inequality holds $A \left( \eta' \Psi_T + \frac{\eta'}{N - A - 2} \Psi_Q \right) + (N - A) < \eta^2,
$
then the modified Krum function offers improved robustness over the original Krum function, i.e., $
\alpha' < \alpha.$
\end{corollary}

\section{Experimental Results}
\begin{figure}[h]
    \centering
    \begin{subfigure}{0.48\linewidth}
        \centering
        \includegraphics[height=2.25cm]{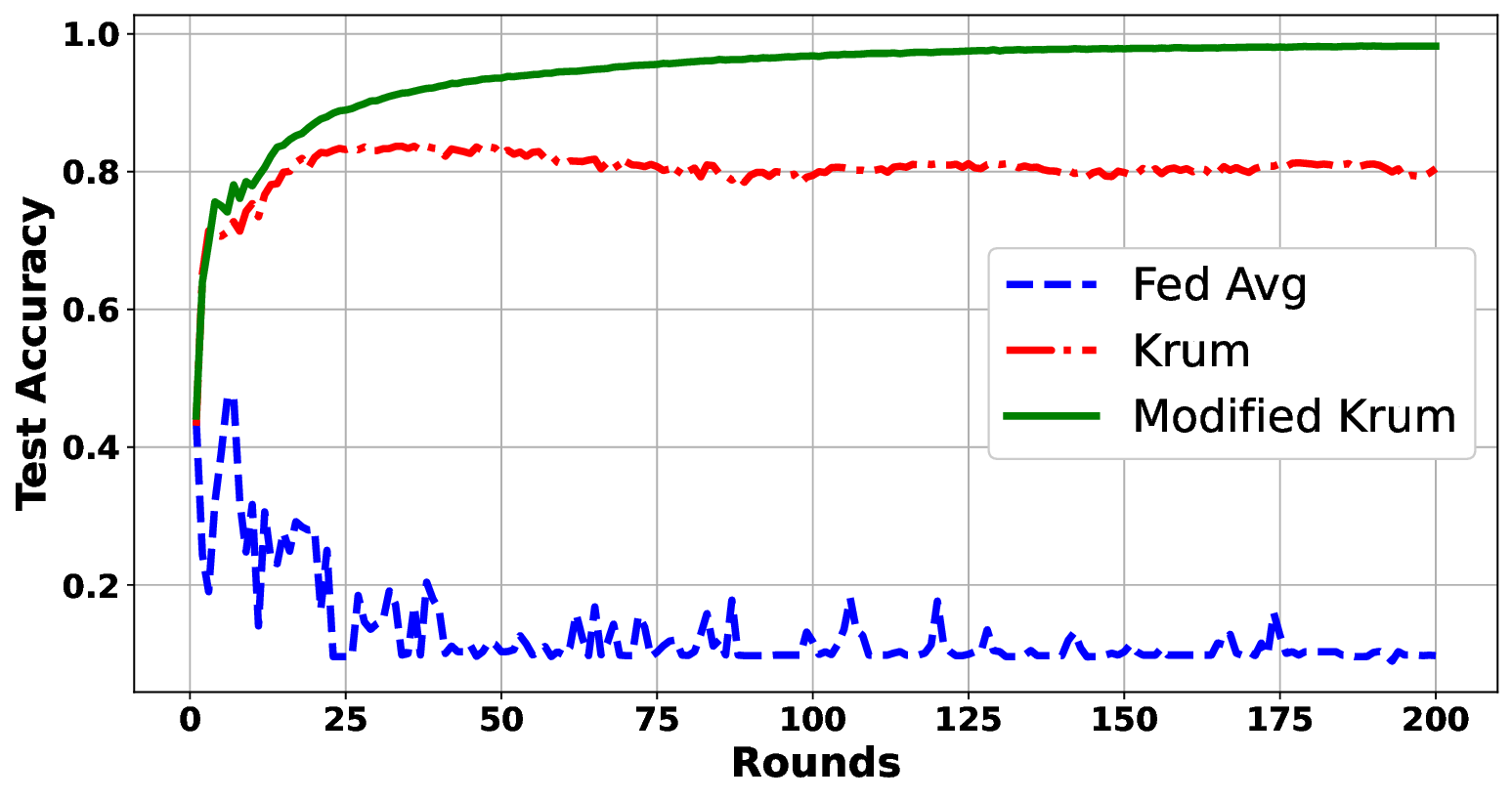}
        \caption{MNIST}
        \label{fig:mnist_results_mnist}
    \end{subfigure}
    \hfill
    \begin{subfigure}{0.48\linewidth}
        \centering
        \includegraphics[height=2.25cm]{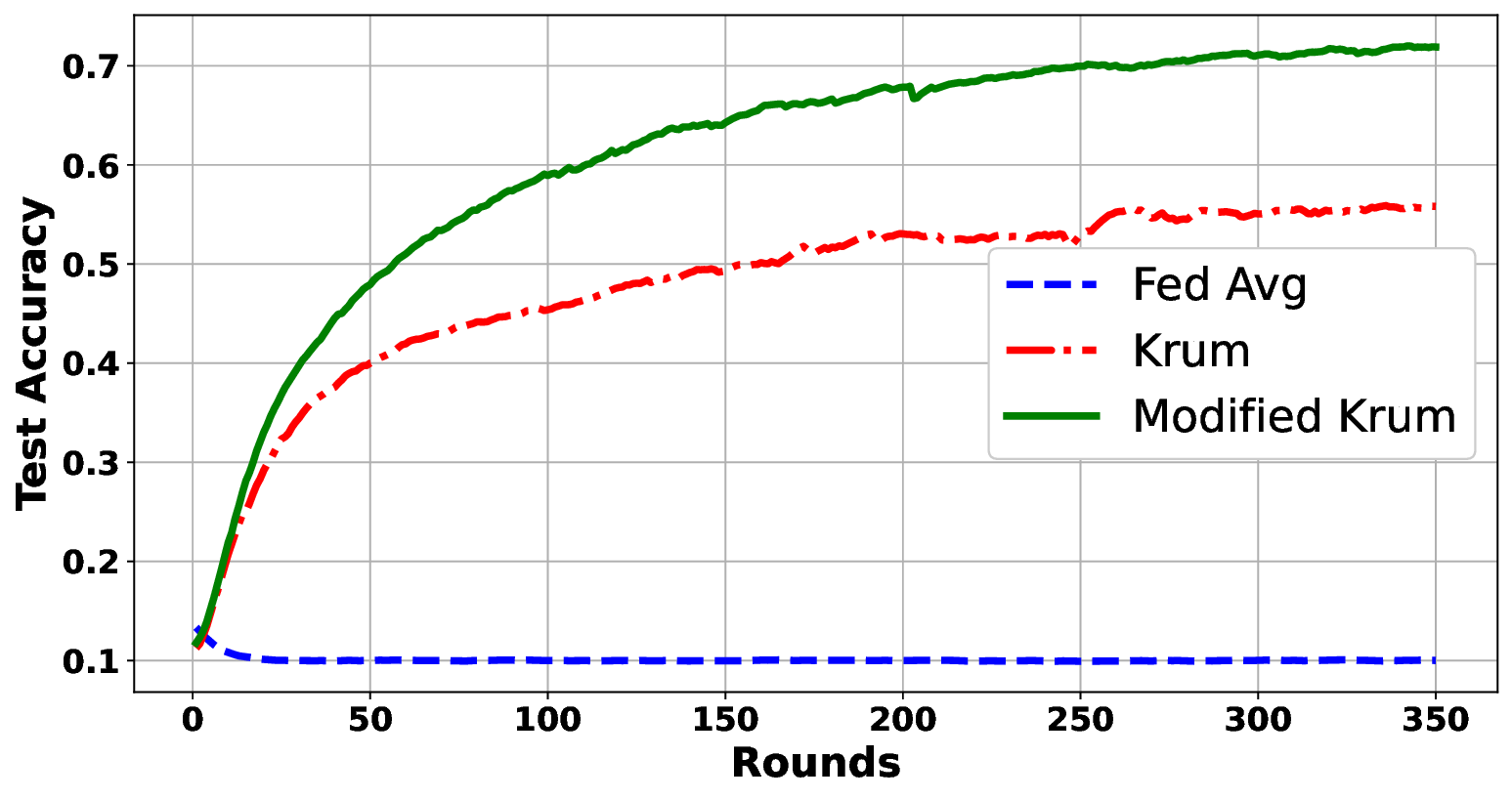}
        \caption{CIFAR-10}
        \label{fig:mnist_results_cifar}
    \end{subfigure}
    \caption{Test accuracy over training rounds on i.i.d. MNIST and CIFAR-10 datasets, comparing FedAvg, Krum, and Modified Krum under Byzantine attacks.}
    \label{fig:test_accuracy_comparison}
\end{figure}

We evaluate the robustness of modified Krum against FedAvg and the original Krum under Byzantine attacks. FedAvg serves as a baseline without outlier detection, making it highly vulnerable to adversarial updates. The original Krum filters outliers solely based on pairwise distances, while our modified Krum incorporates feedback from the DFT decoder to enhance Byzantine detection.
Experiments are conducted on i.i.d. MNIST and CIFAR-10 datasets using \(N = 30\) users, with \(T = 9\) colluding users and \(A = 10\) Byzantine users. Each user trains locally for a few epochs before aggregation. Byzantine users launch model poisoning attacks, adding noise or scaling updates, and inject noise into secret shares to corrupt pairwise distances, as outlined in the threat model.
We select \(m = 8\) updates per round for aggregation. Figures~\ref{fig:mnist_results_mnist} and~\ref{fig:mnist_results_cifar} show the test accuracy over training rounds. As expected, FedAvg performs poorly under attack. Original Krum offers better performance than FedAvg but suffers in test accuracy when compared to modified Krum, validating the benefits of DFT-guided Krum.

      
\section{Conclusion}
In this paper, we introduced  FORTA, a real‐domain secure aggregation framework for FL that simultaneously ensures  privacy and Byzantine resilience. By employing analog secret sharing, FORTA preserves the fidelity of real‐valued model updates without quantization. To defend against finely‐crafted precision‐aware attacks, we integrated decoder‐derived feedback into a modified Krum aggregator, dynamically reweighting user contributions based on estimated adversarial likelihood. Our theoretical analysis and experiments on MNIST and CIFAR‑10 demonstrated that FORTA significantly outperforms standard Krum and FedAvg under Byzantine threats. 

Future work will focus on reducing the computational and communication overhead of analog sharing, as well as establishing theoretical guarantees for convergence for the modified Krum under adversarial conditions.

\bibliographystyle{ieeetr}
\bibliography{references} 

\end{document}